**COVID-19 infection and recovery in various countries: Modeling the dynamics and evaluating the non-pharmaceutical mitigation scenarios**

Yong Zhang [1], Xiangnan Yu [2,*], HongGuang Sun [2], Geoffrey R. Tick [1], Wei Wei [3], and Bin Jin [4,*]

[1] Department of Geological Sciences, University of Alabama, Tuscaloosa, AL 35487, USA

[2] College of Mechanics and Materials, Hohai University, Nanjing 210098, China

[3] Jiangsu Provincial Key Laboratory of Materials Cycling and Pollution Control, Jiangsu Engineering Laboratory of Water and Soil Eco-remediation, School of Environment, Nanjing Normal University, Nanjing 210023, China

[4] Department of Dermatology, Maternal and Child Health Care Hospital of Qixia District, The Second Affiliated Hospital of Nanjing Medical University, Nanjing 210028, China

* Correspondence: yuxiangnan931019@foxmail.com; myjinbin@163.com

**Abstract:** The coronavirus disease 2019 (COVID-19) pandemic radically impacts our lives, while the transmission/infection and recovery dynamics of COVID-19 remain obscure. A time-dependent Susceptible, Exposed, Infectious, and Recovered (SEIR) model was proposed and applied to fit and then predict the time series of COVID-19 evolution observed in the last three months (till 3/22/2020) in various provinces and metropolises in China. The model results revealed the space dependent transmission/infection rate and the significant spatiotemporal variation in the recovery rate, likely due to the continuous improvement of screening techniques and public hospital systems, as well as full city lockdowns in China. The validated SEIR model was then applied to predict COVID-19 evolution in United States, Italy, Japan, and South Korea which have responded differently to monitoring and mitigating COVID-19 so far, although these predictions contain high uncertainty due to the intrinsic change of the maximum infected population and the infection/recovery rates within the different countries. In addition, a stochastic model based on the random walk particle tracking scheme, analogous to a mixing-limited bimolecular reaction model, was developed to evaluate non-pharmaceutical strategies to mitigate COVID-19 spread. Preliminary tests using the stochastic model showed that self-quarantine may not be as efficient as strict social distancing in slowing COVID-19 spread, if not all of the infected people can be promptly diagnosed and quarantined.

# 1. Introduction

The novel coronavirus disease 2019 (COVID-19) outbreak, a respiratory illness that started (first detected) in late December 2019, is a pandemic infecting >336,000 people in more than 140 countries with the average fatality rate of 4.4% globally (data up to 3/22/2020) [1]. The COVID-19 pandemic is infiltrating almost every aspect of life, damaging global economy and altering both man-made and natural environments. Urgent actions have been taken but further effective and efficient strategies are promptly needed to confront this global challenge. To address this challenge and promptly guide the next efforts, it is critical to model the transmission, infection, and recovery dynamics of COVID-19 pandemic. Mathematical models are among the necessary tools to quantify the COVID-19 dynamics and are the primary objective motivating this study.

This study aims to model the COVID-19 evolution for representative countries with apparent coronavirus cases, including China, United States (U.S.), Italy, Japan, and South Korea. COVID-19 spread in these countries under different starting (initiation and detection) times. For example, China had passed the peak of the coronavirus outbreak and hit a milestone with no new local infections on 3/19/2020 (79 days from the onset, 12/31/2019, in Wuhan, China), while the U.S. coronavirus cases soared past 10,000 on the same day. This study will apply the core characteristics of COVID-19 outbreak obtained in China to estimate the COVD-19 spread in the U.S., as well as other countries where the number of affected people has not yet reached its peak.

In addition, this new pandemic may last for a relatively longer time than expected [2]. No vaccine against SARS-CoV-2 (severe acute respiratory syndrome coronavirus 2) is currently available [3]. Indeed, a vaccine for prevention and infection control may not be ready before March 2021 for COVID-19, considering the minimum 4~5 weeks for trial and at least 1 year for safety evaluation and final deployment. Efficient strategies are therefore needed to mitigate the COVID-19 outbreak. Possible non-pharmaceutical scenarios, such as isolation of cases and contact tracing, can be evaluated using mathematical models [4], to identify the most efficient strategy going forward. This is another major motivation and the secondary task of this study.

To address the questions mentioned above, this study is organized as follows. Section 2 proposes an updated SEIR model for COVID-19, where "S", "E", "I", and "R" stand for Susceptible, Exposed, Infectious, and Recovered people, respectively [5]. This model is then applied to fit and predict the COVID-19 spread in various provinces and major cities in China, resulting in abundant datasets to derive the core characteristics of the COVID-19 dynamics of transmission/infection and recovery. Section 3 predicts the spread of COVID-19 in the other countries using the knowledge gained from China. Section 4 proposes a fully Lagrangian approach to model the spatiotemporal evolution of COVID-19, and then applies it to evaluate non-pharmaceutical scenarios to mitigate the virus spread. Section 5 reports the main conclusions.

## 2. A time-dependent SEIR model for the COVID-19 spread in China

Several mathematical models have been applied for epidemic analysis of COVID-19. The most widely used one, so far, is the well-known SEIR model. For example, Peng et al. [6] (using data up to 2/16/2020) proposed a generalized SEIR model to successfully estimate the key epidemic parameters of COVID-19 in China, and predicted the inflection point and ending time of confirmed COVID-19 cases. The SEIR model was also applied by Li et al. [7] (using the observed data up to 2/6/2020) to compare the effect of city lockdowns on the transmission dynamics in different cities in China. The SIR model with time-dependent transmission and recovering rates was used by Chen et al. [8] (using data up to 2/20/2020) to analyze and predict the number of confirmed cases of COVID-19 in China. The SIR model was extended by Wang et al. [9] (using data up to 2/12/2020) to incorporate various time-varying quarantine protocols for assessing interventions on the COVID-19 epidemic in China. The SEIR model and its modifications were also successfully applied by others [10-13], mostly for assessing the early spreading of COVID-19 in China. Previous applications of the popular SEIR model, however, may contain high uncertainty since they had limited data access for only a short period of the COVID-19 outbreak. As will be shown below, the COVID-19 dynamics have changed dramatically in the last three months, likely due to the improvement/adjustment of screening/testing techniques, public hospital system capabilities, and the government's control policies for contagious diseases. An updated, much better version of the SEIR model is therefore needed and can be reliably built now for China, since the detailed datasets are now available and the coronavirus outbreak (i.e., # of people infected) has passed its peak in China.

Other models have also been used for specific purposes related to COVID-19, including the global metapopulation disease transmission model to project the impact of travel limitations on the epidemic spread [14], the transmission model for risk assessment [15], and the synthetic contact matrix model for reproductive ratios for COVID-19 [16]. To the best of our knowledge, the classical stochastic epidemic models, such as the discrete or continuous time Markov chain model and the stochastic differential equation model, have not yet been applied for COVID-19 spread scenarios. A preliminary stochastic model for evaluating COVID-19 spread in a city will be developed and applied in Section 4.

This section focuses on the deterministic model. The classical SEIR model considers constant parameters, which may not well capture time-dependent dynamics. Hence, the SEIR model is updated in Section 2.1., and then tested in Section 2.2.

*2.1. SEIR model update*

The classical SEIR model containing four populations (S, E, I, and R) takes the form [17]:

$$\frac{dS}{dt} = -rp\,I\frac{S}{N} \qquad (1a)$$

$$\frac{dE}{dt} = rp\,I\frac{S}{N} - \alpha E \qquad (1b)$$

$$\frac{dI}{dt} = \alpha\,E - \gamma\,I \qquad (1c)$$

$$\frac{dR}{dt} = \gamma\,I \qquad (1d)$$

where $S$ is the stock of susceptible population; $E$ is the number of persons exposed to or in the latent period of the disease; $I$ is the stock of infected; $R$ is the stock of recovered population; $r$ denotes the number of susceptible people whom the infected people contact daily; $N$ is the sum of all the four groups of people ($S(t) + E(t) + I(t) + R(t) = N$ =constant representing the constancy of population $N$); $p$ is the constant rate of infection (i.e., representing the probability for the infected people to transform the susceptible people into infected ones); $\alpha$ is the constant rate for the exposed person transformed into the infected one; and $\gamma$ is the constant recovery rate (defining the speed for the infected person to be cured or expired).

To allow for possible time-sensitive rates for COVID-19 evolution, we revise model (1):

$$\frac{dS}{dt} = -r\,p(t)\,I\frac{S}{N} - r_2\,p_2(t)\,E\frac{S}{N} \qquad (2a)$$

$$\frac{dE}{dt} = r\,p(t)\,I\frac{S}{N} - \alpha E + r_2\,p_2(t)\,E\frac{S}{N} \qquad (2b)$$

$$\frac{dI}{dt} = \alpha\,E - \gamma(t)\,I \qquad (2c)$$

$$\frac{dR}{dt} = \gamma(t)\,I \qquad (2d)$$

$$\frac{d^\beta D}{dt^\beta} = \gamma_\beta\,I \qquad (2e)$$

where $D$ represents the number of deaths (which is one component in $I$); $p_2$ is the rate for the healthy, susceptible person to be transferred to the infected one by the exposed people (note that COVID-19 patients in the incubation period might be contagious too); and $r_2$ is the number of healthy susceptible people that are contacted by the exposed people daily. Now, the infection rates can change with time, and the infected persons are removed from the risk of infection also with a time-dependent rate of $\gamma(t)$. If the recovered individuals can return to the susceptible status due to for example loss of immunity, then the partial differential equation (PDE) (2d) for the time rate of change of $R$ needs one more (sink) term: $dR/dt = \gamma(t) I - \xi R$, where $\xi$ is the rate of the recovered individuals returning to the susceptible status.

We add the fractional-order PDE (2e) containing the death probability of $\gamma_\beta$ (while the other patients are cured): $\frac{d^\beta D}{dt^\beta} = \frac{1}{\Gamma(1-\beta)} \int_0^t \frac{\partial D(\tau)}{\partial \tau} (t-\tau)^{-\beta} d\tau$, which is the Caputo fractional derivative [18, 19] with order $\beta$ ($0 < \beta \leq 1$). When the order $\beta=1$, model (2e) reduces to the classical integer-order PDE for the death evolution. The fractional PDE (2e) is used here for two reasons. First, the evolution of deaths and cures may be characterized by a random process, since the exact time for the (recovered) person to be initially infected is unknown (i.e., the patient that died or was cured today may have been diagnosed yesterday or last week). Second, some patients may not be treated in time after being infected, making the death toll to evolve with a time memory. Therefore, we extend the classical mass-balance equation of death cases to the fractional PDE to characterize the random property and memory impact embedded in the temporal evolution of mortality. The fractional PDE (2e) and its classical version will be compared below using real data.

*2.2. Result analysis for province and major cities in China*

We apply the SEIR model (2) to fit the infection and recovery of coronavirus in China up to 3/7/2020, and then predict the future evolution (**Figure 1**). The measured time series data from 3/8/2020 to 3/22/2020 are used to check the model predictions. The epicenter of the infected population has been under full lockdown to limit the spread of coronavirus, and hence the infection dynamics of coronavirus in China represent one of the strict mitigation scenarios of coronavirus.

2.2.1. Comparison of different models

Other complex SEIR models were also applied to model COVID-19 spread, such as the one proposed by Tang et al. [20] which has 12 rates/probabilities and 8 groups of people. Numerical results show that, compared with the SEIR model (2), the complex model proposed by Tang et al. [20] (with solutions shown by the dotted lines in **Figure 1a**) accurately fits the observed data at the early stage, but then overpredicts the spread of COVID-19 observed after 2/12/2020. Dynamics of transmission for COVID-19, especially the recovery rate, therefore, changed in time in China, likely due to the time-dependent conditions in for example medical care mentioned above. A dynamic SEIR model, therefore, may be preferred for modeling COVID-19 spread in China.

In addition, compared to the fractional PDE (2e), the best-fit solution using the classical PDE $dD/dt = \gamma_\beta I$ for death evolution (see the black, dotted line in **Figure 1a**) slightly overestimates the late-time growth of mortality. The actual death toll in Hubei province grew slower than that estimated by a constant rate model, indicating that the memory impact may affect the late-time dynamics of death that can be better captured by the fractional FDE (2e).

2.2.2. Comparison of different regions in China

The best-fit solutions using model (2) fit the evolution of the infected and recovered populations well for the data recorded from Hubei province and three large cities closely related to Wuhan, China (**Figure 1**). The model can also predict well the observed time series of COVID-19 spread from 3/8/2020 to 3/22/2020 for most places, except for Shanghai City (**Figure 1d**). This is due to the oversea cases imported to Shanghai, whose number was increasing quickly after 3/3/2020, causing inconsistency of population and failure of the model. Shanghai Pudong International Airport, one of the two airports located in Shanghai City, is the eighth-busiest airport in the world and the busiest international gateway of mainland China. If deleting the coronavirus cases imported from overseas, model (2) can predict well the COVID-19 data in Shanghai (**Figure 2**). Therefore, model (2) works well for various places in China, while external sources can easily break the internal evolution, especially the asymptotic status, of COVID-19 in China.

The resultant time-dependent recovery rate $\gamma(t)$ is depicted in **Figure 3**, where the rate fitted by the latest observation data point in the fitting period (i.e., 3/7/2020) remains stable in the following prediction period. The best-fit recovery rate is the highest for Shanghai (except for the impulse of $\gamma(t)$ for Wenzhou discussed below), which is expected since Shanghai has the best public health system of all of these cities. Contrarily, Hubei shows the lowest recovery rate, likely due to its delayed response and the relatively limited public health capability at the beginning of the outbreak compared with Shanghai.

Wenzhou exhibits an impulse in the infection and recovery dynamics of coronavirus (**Figure 1b**), different from the other places. On February 27, 2020, the number of Wenzhou's infected people suddenly declined, combined with the sudden increase of the number of people cured. This abrupt change can be efficiently captured by the SEIR model (2) with an impulse in the recovery rate $\gamma(t)$. This impulse is most likely due to the new hospital, the No. 2 Affiliated Hospital of Wenzhou Medical University, built in this city in early February, which significantly improved the public health system. The first discharged cases of coronavirus from this hospital appeared in late February 2020, resulting in the sudden increase in the total recovery rate. In addition, a relatively large number of people working in Wuhan returned to Wenzhou in late January, and it appears that the improved efficient screening process successfully identified the number of infected cases. The new cases were ~29,000 from 1/24/2020 to 1/31/2020 in Wenzhou (with an average of 3,600 new patients per day), who were then immediately centralized for treatment. It appears that this fast response helped to alleviate the spread of coronavirus in Wenzhou.

2.2.3. Parameter analysis

The best-fit parameters of model (2) are listed in **Table 1**, and the initial values for each group of people are listed in **Table 2**. We reveal three behaviors in model parameters. First, the best-fit "S"-shaped $\gamma(t)$ (**Figure 3**) can be described by the sigmoid function $\gamma(t) = a/(1 + e^{bt+c})$ (where $a$, $b$, and $c$ are factors), showing that the recovery rate increases exponentially before reaching stable. This increase is likely due to the healthcare facility and experience improved with time (in an accelerating rate) before reaching their asymptote or maximum capacity.

Second, the rates and probabilities ($r$, $r_2$, $p$, and $p_2$) affecting the COVID-19 transmission/infection slightly change in space and remain stable for a given site (**Table 1**). The small spatial fluctuation of these rates may be due to the similar strategies of local government for intervening COVID-19 (i.e., full lockdowns) across the country. The time-invariant transmission/infection rates for a given site imply that the spreading rate of COVID-19 may not significantly change in a short time for a given area with a consistent intervening policy.

Third, the model predictions show that it may take 47, 14, 25, and 11 days from 3/14/2020 for the current infected population to drop below 1% of the cumulative infected population for Hubei province, and the cities of Wenzhou, Shenzhen, and Shanghai, respectively. Notably, since the best-fit $\gamma(t)$ increases in time, the prediction using a constant $\gamma$ may overestimate the long-term future infection and underestimate the future recovery of coronavirus.

We also introduce an index $C$ to quantify the infection severity of COVID-19 at different places:

$$C = \frac{N}{I_{max}} \quad (3)$$

where $I_{max}$ is the maximum number of cumulative infected people at the given site. A smaller $C$ represents a greater infection severity of coronavirus. There is a power-law relationship between the regional population $N$ and the maximum cumulated number of infected people $I_{max}$ (**Figure 4**). This empirical formula may be used to approximate the largest cumulative number of infections, which will be applied below in predicting the COVID-19 evolution outside of China where the coronavirus infection has not yet reached its peak number of cases.

**3. Prediction of spreading of COVID-19 in the world**

Different countries are applying different modes for slow the COVID-19 spread. In the next sub-sections, we discuss several representative countries and then fit/predict the virus spread there.

*3.1. United States*

To decrease the acceleration of COVID-19 spread, the U.S.'s main propagated mode up to 3/22/2020 has been the use of social distancing actions. The predicted COVID-19 spread in the U.S. is plotted in **Figure 5**. There is a very high uncertainty in the prediction of COVID-19 evolution of in the U.S., considering the recent dramatic fluctuations in the number of infected people and the delayed screening during the beginning stage of the outbreak. Thus, we must make assumptions and can only estimate the future coronavirus evolution in the U.S. under specific scenarios. Based on the index of infection severity (*C*) in the representative areas of China, we predict the U.S. COVID-19 evolution dynamics under two scenarios as shown in **Figure 5**.

**Figure 5a** shows the possible spread of COVID-19 in California, U.S., assuming that the coronavirus dynamics (including the infection/recovery rates and the infection severity) are similar to those observed in Shanghai, China (the solid lines, representing an optimistic scenario) or Hubei, China (the dashed lines, representing a "normal" scenario). The estimated total population of infection in California vary from 5,000 to 45,000, with the peak number of cases appearing around ~10 to ~30 days from 3/22/2020.

**Figure 5b** shows the possible spread of COVID-19 in the U.S. Solid lines are solutions assuming that COVID-19 spreads similarly as that in Shanghai, China, while the dashed lines are solutions assuming COVID-19 spreading similarly as that in Hubei, China. The estimated total population of infection vary from 110,000 to 740,000, with the peak case number appearing 10~31 days from 3/22/2020 and the COVID-19 outbreak lasting for 35~80 days from 3/22/2020.

*3.2. Italy*

To decrease the acceleration of COVID-19 spread, Italy's mode is now similar to China: lockdown of the full population. The predicted COVID-19 spread in Italy is plotted in **Figure 6a**. Although Italy has followed China's mode of national isolation, the number of infected people increased rapidly from 3/14/2020 to 3/19/2020 (~495 new cases per day). To account for the delayed national quarantine compared with China, we decrease the *C* index (when also increasing the upper limit of the cumulative infection). The COVID-19 evolution prediction results show that there may be a turning point in the next two weeks when the current infected cases begin to decline. We also separate the death toll from the number of recovered cases.

*3.3. South Korea*

South Korea's mode of combating the spread of COVID-19 is fast detection and tracking of the disease. South Korea is using efficient mobile diagnoses tests and accurate tracing of infected cases, to maintain a low death rate even with a large infected population. The mobile method can test 20,000 people per day (the maximum capability on 3/12/2020), and apps for cell phones and/or credit cards can accurately track the routes of infected people with the help of local government (without invasion of privacy), so that warnings can be immediately delivered to the general population to obviate the places with high risk. The current infected population may have passed its peak number of cases around 3/20/2020, and the prediction shows that the COVID-19 outbreak may be well controlled in ~35 days from 3/22/2020 (**Figure 6b**).

*3.4. Japan*

Japan's mode of combating the spread of COVID-19 is compatible with that of the U.S., in addition to other changes such as enhanced education/outreach and fast treatment to the infected cases. Specific policies include social distancing (which might be a key barrier to the spread of the novel coronavirus), personal hygiene, and quarantine of the infected cases. The current data and modeling results (**Figure 6c**) show that Japan has an efficient way so far to limit the maximum population infected and slow the spread of COVID-19, while this outbreak may last for a while.

**4. Stochastic model to evaluate scenarios to mitigate the outbreak of COVID-19 in a city**

The model-predicted COVID-19 spread in U.S., using the fitted infection and recovery rates from China (**Figure 5**), reveals the impact of one possible mitigation scenario for COVID-19: state coronavirus lockdowns, which have now been implemented by some states in the U.S. such as California and New York. Other non-pharmaceutical options can and should also be evaluated using mathematical models, considering the recent surge of infected cases in the U.S.

When the number of infected persons is initially small compared to that of susceptible people, the infected and susceptible people are not well mixed and hence the system is not homogeneous. Under such conditions, a stochastic model is needed, as the deterministic, continuum models (such as the SEIR model) assume well-mixing of components for a homogeneous system [21-22]. Hence, this section develops and applies a stochastic model to evaluate the non-pharmaceutical scenarios for mitigating COVID-19 with a small number of initial infections.

*4.1 Stochastic model development*

The random walk based stochastic model for COVID-19 spread is analogous to a mixing-limited bimolecular reaction-based mechanism/condition [23]. When a reactant *A* particle (representing a susceptible individual $A$) meets a reactant $B$ particle (representing an infectious person $B$), a chemical reaction may occur if the collision energy is large enough to break the chemical bond (meaning that the susceptible person $A$ may be infected if satisfying additional criteria such as $A$ and $B$ are close enough, and $B$ touches his/her face after receiving coronavirus from $A$). Therefore, the condition of $A$ being infected is not deterministic but rather a random, probability-controlled process. This probability is related to various factors, such as the duration that $A$ and $B$ are in contact, the infectivity rate, and the distance between the two people, which may be characterized parsimoniously by the interaction radius $R$ that controls the number of reactant pairs (susceptible + infectious) in a potential reaction (infection) [23]. Hence, the core of the random walk stochastic model for COVID-19 spread is to define the interaction radius $R$.

The analogous development and similarities between bimolecular reactions and the SIR model can also be seen from their governing equations. The time-dependent SIR model takes the form [24]:

$$\frac{d\,S(t)}{dt} = \frac{-\beta(t)\,S(t)\,I(t)}{n} \qquad (4a)$$

$$\frac{d\,I(t)}{dt} = \frac{\beta(t)S(t)\,I(t)}{n} - \gamma(t)I(t) \qquad (4b)$$

$$\frac{d\,R(t)}{dt} = \gamma(t)I(t) \qquad (4c)$$

$$S(t) + I(t) + R(t) = n \qquad (4d)$$

where $\beta(t)$ and $\gamma(t)$ denote the transmission rate and recovery rate at time $t$, respectively.

The advection-dispersion-reaction (ADR) equation for irreversible bimolecular reaction $A + B \rightarrow C$ takes the form [23]:

$$\frac{d\,[A(x,t)]}{dt} = -v_A \frac{\partial\,[A(x,t)]}{\partial x} + D_A \frac{\partial^2\,[A(x,t)]}{\partial x^2} - K_f(t)[A(x,t)][B(x,t)] \qquad (5a)$$

$$\frac{d\,[B(x,t)]}{dt} = -v_B \frac{\partial\,[B(x,t)]}{\partial x} + D_B \frac{\partial^2\,[A(x,t)]}{\partial x^2} - K_f(t)[A(x,t)][B(x,t)] \qquad (5b)$$

$$\frac{d\,[C(x,t)]}{dt} = -v_C \frac{\partial\,[C(x,t)]}{\partial x} + D_C \frac{\partial^2\,[C(x,t)]}{\partial x^2} + K_f(t)[A(t)][B(t)] \qquad (5c)$$

where $[A]$, $[B]$, and $[C]$ denote the concentrations of $A$, $B$, and $C$, respectively; $v_A$, $v_B$ and $v_C$ denote the mean flow velocity for $A$, $B$, and $C$, respectively; $D_A$, $D_B$ and $D_C$ denote the macrodispersion coefficient for $A$, $B$, and $C$, respectively; and $K_f(t)$ is the forward kinetic coefficient of reaction. Equations (5a) and (5b), when simplified to the rate equations, are functionally similar to equations (4a) and (4b), respectively, if the recovery rate $\gamma(t) = 0$. Therefore, following the argument in Zhang et al. [23] and Lu et al. [25], we derive analytically the interaction radius $R$ for the SIR model (4):

$$R(t) = \left[\frac{1}{2\pi} \left|\frac{\beta(t)}{n} \frac{m_A}{V} \Delta t - \gamma(t)\Delta t\right|\right]^{1/2}, \qquad (6)$$

where $V$ denotes the volume of the domain, $\Delta t$ is the time step in random walk particle tracking, $m_A = [A_0]V/N_A^0$ is the mass (or weight) carried by each $A$ particle, $[A_0]$ is the initial concentration of $A$ (which can be assumed to be the normalized value 1 here), and $N_A^0$ denotes the initial number of susceptible people. After defining the interaction radius $R$, the particle tracking scheme proposed by Zhang et al. [23] and Lu et al. [25] can be applied to model the transmission of coronavirus between the susceptible and infectious people.

*4.2 Scenario evaluation to mitigate the COVID-19 outbreak in a city*

In addition to pharmaceutical strategies including vaccine and therapeutic drug development, and herd immunity that may either take a while or have a high risk, non-pharmaceutical scenarios can be tested. Several particle-tracking based stochastic models were proposed recently [26] to evaluate non-pharmaceutical scenarios to mitigate coronavirus spreading in a city. Here we evaluate three related scenarios (described below) using the stochastic model proposed above.

*Scenario 1*: No special constraints. Assuming a city with 10,000 people and 4 initial coronavirus cases. This scenario does not put any constraints for any population. We assume that the population distributes in the city randomly at the beginning, and then moves randomly. The spread of COVID-19 is then simulated using the stochastic model proposed in Section 4.1.

*Scenario 2*: Social distancing. This scenario assumes that all people in the city maintain a social distance, significantly decreasing the probability of infection.

*Scenario 3*: Attempted quarantine. This scenario creates a forced quarantine for infected individuals before the outbreak (which is expected to be the most efficient way of quarantine).

In the stochastic model, we assume that 10 days after being infected, the person will be removed because of being cured or expired (dead). This is because the median disease incubation period was estimated to be 5.1 days [27]. For simplicity purposes, the interaction radius $R$ (6) remains constant, since the constant interaction radius was found to be able to efficiently capture the temporal variation of effective reaction rates in mixing controlled reactions [23, 25]. The initial number of $A$ and particles is 10,000 and 4, respectively. The Lagrangian solutions of the COVID-19 outbreak for the three scenarios are depicted in **Figure 7**.

Scenarios 1, 2, and 3 have a peak in the curve of newly infected people at time $t$=28, 65, and 32, respectively, showing that the virus spreads the fastest for the scenario without mitigation constraints (i.e., scenario 1, where the number of the total infected increases by one order of magnitude every 10 days in the rising limb), as expected. However, the value for this peak (scenario 1, =198 people) is lower than that for scenario 3 (=267 people), although the total number of the infected people for scenario 1 (9,336) is slightly larger than scenario 3 (9,322). This may be due to a greater separation of infection cases for the higher number of initial coronavirus carriers in scenario 1, which causes a lower and relatively flatter COVID-19 evolution peak than scenario 3. Scenario 2 has the lowest peak value (=121 people) and the most-delayed peak in the curve of new cases, and the total infection time is almost doubled compared to the other two scenarios, indicating that people living with strict social distancing may also suffer from a much longer period of COVID-19 threat. It is also noteworthy that the overall trend of the solution of scenario 1 (initial surge without special constraints, **Figure 7a**) is similar to that for Italy which had delayed response to the COVID-19 outbreak initially (**Figure 6a**), and scenario 3 solution (a lower peak value and a longer duration due to social distancing, **Figure 7b**) is similar to that for Japan which has been taken social distancing actions (**Figure 6c**).

The simulated particle plumes plotted in **Figure 8** reveal the subtle discrepancy between the three mitigation scenarios. *Scenario 1* assumes that four initial cases were initially located on the right side of the city, while the whole population (10,000 susceptible persons) was distributed randomly in the 1×1 domain (**Figure 8a**). The trajectory of each person is assumed to follow (two-dimensional) Brownian motion with retention, to capture the random vector for each displacement and the random waiting time between two consecutive motions. The virus moved quickly from east to west (**Figures 8b** and **8c**), spreading over the whole city before all the infected people were cured or expired (dead) at time $t=69$ (**Figures 8d**). A total of 664 susceptible people (6.6% of the total population) distributed randomly around the city were never infected.

*Scenario 2* assumes that social distancing can reduce the infection probability, which can be characterized by a smaller reaction rate or a smaller interaction radius in our Lagrangian approach. The virus was spreading at a much lower rate from west to east than that in scenario 1 (**Figures 8e~8g**), reaching stable (i.e., # of cases) at an apparently later time ($t=125$) and leaving more susceptible people unaffected (2,734 total, which is 27.3% of the population). Therefore, social distancing is effective in limiting the spread of coronavirus among people. Note, however, this scenario assumes that every person in this city strictly maintains social distancing; otherwise a surge of infections may occur the same way as that shown in scenario 1.

*Scenario 3* assumes self-quarantine. Notably, not all of the infected people can be effectively quarantined due to the following: 1) people can be infected without coronavirus symptoms; 2) people in the incubation period can transmit the infection; and 3) limited health care facilities for the large influx of patients. For example, according to Imai et al. [28], many infected people could not be appropriately screened initially in Wuhan City, China. Under this condition, we assume that 50% of people infected and diagnosed are immediately quarantined, while the remaining infected people (**Figure 8i**) can still cause the spread of coronavirus (**Figures 8j~8l**). Self-quarantine, therefore, may not be as effective as maintained social distancing.

## 5. Conclusions

The COVID-19 pandemic radically impacts our lives, altering our daily patterns and interactions in an unprecedented way and rate. In this study, the transmission/infection and recovery dynamics of COVID-19 were quantified computationally through well-established analysis of PDEs that describe transmission/infection and recovery impacts through time. A time-dependent SEIR model, motivated by the coronavirus milestone of recording zero new local infections in China on 3/19/2020, was applied to capture the COVID-19 evolution observed in the last three months (up to 3/22/2020) in China. A stochastic model was also proposed to evaluate non-pharmaceutical mitigation options. Four main conclusions are presented by this study.

First, the transmission/infection rate of COVID-19 changes slightly between cities in China, while the recovery rate of COVID-19 increases apparently with time due to various human factors. The revised SEIR model revealed a significant time variation (either gradually or abruptly) in the recovery rate which also changes in locations in China. For example, a notable increase of COVID-19 recovery was identified for the city of Wenzhou, China (whose population has a direct connection to Wuhan), likely due to the dramatic improvement of the local public hospital system.

Second, the model results showed that, if the infection and recovery rates remain stable, then the number of cumulative infections is likely to decline to 1% of the total infected population within one month (from 3/22/2020) in China. The overseas imported cases, however, may break the internal asymptotic status and significantly postpone the ending time of COVID-19 in China.

Third, predictions using the validated SEIR model showed different evolution dynamics of COVID-19 within populations of the United States, Italy, Japan, and South Korea. For example, it may take at least 1~3 months for the current infected population to drop below 1% of the cumulative infected population in the U.S. (requiring more efficient mitigation strategies), while South Korea may control the COVID-19 spread in less than 35 days from 3/22/20. However, these predictions contain high uncertainty due to uncertainty in the maximum infected population, and changes of transmission/infection and recovery rates among different countries.

Fourth, a stochastic model based on the Lagrangian scheme, analogous to a mixing-limited reaction mechanism model, showed that self-quarantine may not be as effective as strict social distancing, since not all the infected people can be diagnosed and immediately quarantined. While strict social distancing can apparently slow COVID-19 spread, the pandemic may last longer.

# References

1. Johns Hopkins Coronavirus Resource Center, **2020**. https://coronavirus.jhu.edu/map.html

2. Yang, C.; Wang, J. A mathematical model for the novel coronavirus epidemic in Wuhan, China. *Math. Biosci. Eng.* **2020**, *17*, 2708–2724.

3. del Rio, C.; Malani P.N. COVID-19 – New insights on a rapidly changing epidemic. *JAMA* **2020**, doi:10.1001/jama.2020.3072

4. Hellewell, J.; Abbott, S.; Gimma, A.; Bosse, N.I.; Jarvis, C. I.; Russell, T.W.; Munday, J.D.; Kucharski, A.J.; Edmunds, W.J. Feasibility of controlling COVID-19 outbreaks by isolation of cases and contacts. *Lancet Glob. Health* **2020**, *8*, e488-e496.

5. Kermack, W.O.; McKendrick, A.G. A contribution to the mathematical theory of epidemics. *P. Roy. Soc. A, Math. Phys. Eng. Sci.* **1927**, *115*, 700–721.

6. Peng, L.; Yang, W.; Zhang, D.; Zhuge, C.; Hong, L. Epidemic analysis of COVID-19 in China by dynamical modeling. **2020**, https://arxiv.org/abs/2002.0656

7. Li, X.H.; Zhao, X.; Sun, Y.H. The lockdown of Hubei province causing different transmission dynamics of the novel coronavirus (2019-ncov) in Wuhan and Beijing. medRxiv, **2020**, https://www.medrxiv.org/content/10.1101/2020.02.09.20021477.full.pdf

8. Chen, Y.C.; Lu, P.E.; Chang, C.S. A time-dependent SIR model for COVID-19. **2020**, https://arxiv.org/ftp/arxiv/papers/2003/2003.00122.pdf

9. Wang, L.L.; Zhou, Y.; He, J.; Zhu, B.; Wang, F.; Tang, L.; Eisenberg, M.; Song, P.X.K. An epidemiological forecast model and software assessing interventions on COVID-19 epidemic in China. **2020**, https://www.medrxiv.org/content/10.1101/2020.02.29.20029421v1.full.pdf

10. Pan, J.; Yao, Y.; Liu, Z.; Li, M.; Wang, Y.; Dong, W.; Kan, H.; Wang, W. Effectiveness of control strategies for Coronavirus Disease 2019: a SEIR dynamic modeling study. **2020**. https://www.medrxiv.org/content/medrxiv/early/2020/03/03/2020.02.20.20023572.full.pdf
20

11. Shi, P.P.; Cao, S.; Feng, P. SEIR Transmission dynamics model of 2019 nCoV coronavirus with considering the weak infectious ability and changes in latency duration. **2020**. https://doi.org/10.1101/2020.02.16.20023655

12. Wu, J.T.; Leung, K.; Leung, G.M. Nowcasting and forecasting the potential domestic and international spread of the 2019-nCoV outbreak originating in Wuhan, China: a modelling study. *Lancet* **2020**, *395*, 689–697.

13. Zhou, T.; Liu, Q.; Yang, Z.; Liao, J.; Yang, K.; Bai, W.; Lu, X.; Zhang, W. Preliminary prediction of the basic reproduction number of the Wuhan novel coronavirus 2019-nCoV. *J. Evidence-Based Med.* **2020**, *13*, 3-7.

14. Chinazzi, M.; Davis, J.T.; Ajelli, M.; Gioannini, C.; Litvinova, M.; Merler, S.; Piontti, A.P.; Mu, K.; Rossi, L.; Sun, K.; Viboud, C.; Xiong, X.; Yu, H.; Halloran, M.E.; Longini Jr. I. M.; Vespignani, A. The effect of travel restrictions on the spread of the 2019 novel coronavirus (COVID-19) outbreak. *Science* **2020**, 06 Mar 2020: eaba9757, doi:10.1126/science.aba9757

15. Boldog, P.; Tekeli, T.; Vizi, Z.; Dénes, A.; Bartha, F.A.; Rost, G. Risk assessment of novel coronavirus COVID-19 outbreaks outside China. *J. Clin. Med.* **2020**, *9*, 571.

16. Hilton, J., Keeling, M. J. Estimation of country-level basic reproductive ratios for novel Coronavirus (COVID-19) using synthetic contact matrices. **2020**, https://www.medrxiv.org/content/10.1101/ 2020.02.26.20028167v1

17. Brauer, F.; Castillo-Chavez, C.; Castillo-Chavez, C. *Mathematical Models in Population Biology and Epidemiology*. Vol. 2; Springer: New York, U. S., **2012**; pp. 3–47.

18. Caputo, M. Linear model of dissipation whose q is almost frequency independent-II. *Geophys. J. Int.* **1967**, *13*, 529–539.

19. Zhang, Y.; Yu, X.N.; Li, X.C.; Kelly, J.F.; Sun, H.G.; Zheng, C. M. Impact of absorbing and reflective boundaries on fractional derivative modes: Quantification, evaluation and application. *Adv. Water Resour.* **2019**, *128*, 129–144.

20. Tang, B.; Wang, X.; Li, Q; Bragazzi, N.L.; Tang, S.; Xiao, Y.; Wu, J.H. Estimation of the transmission risk of the 2019-nCoV and its implication for public health interventions. *J. Clin. Med.* **2020**, *9*, 462, doi:10.3390/jcm9020462.

21. Allen, L.J.S. *An introduction to stochastic processes with applications to biology*, 2nd ed.; CPC Press: New York, U. S., **2010**; pp. 45–102.

22. Chalub, F.A.C.C., Souza, M.O. Discrete and continuous SIS epidemic models: A unifying approach. *Ecol. Complex.* **2014**, *18*, 83–95

23. Zhang, Y.; Papelis, C.; Sun, P.; Yu, Z. Evaluation and linking of effective parameters in particle-based models and continuum models for mixing-limited bimolecular reactions. *Water Resour. Res.* **2013**, *49*, 4845–4865.

24. Newman, M. Networks: An Introduction. Oxford University Press, **2010**.

25. Lu, B.Q.; Zhang, Y.; Sun, H.G.; Zheng, C. M. Lagrangian simulation of multi-step and rate-limited chemical reactions in multi-dimensional porous media. *Water Sci. Eng.* **2018**, *11*, 101–113.

26. Stevens, H. Why outbreaks like coronavirus spread exponentially and how to "flatten the curve". *The Washington Post*, March 14, **2020**. https://www.washingtonpost.com/graphics/2020/world/corona-simulator/

27. Stephen, A.L.; Grantz, K.H.; Bi, O.; Jones, F.K.; Zheng, O.; Meredith, H.R.; Azman, A.S.; Reich, N.G.; Lessler, J. The Incubation Period of Coronavirus Disease 2019 (COVID-19) From Publicly Reported Confirmed Cases: Estimation and Application. *Ann. Intern. Med.* **2020**, DOI:10.7326/M20-0504

28. Imai, N.; Dorigatti, I.; Cori, A.; Donnelly, C.; Riley, S.; Ferguson, N.M. Report 2: Estimating the potential total number of novel Coronavirus cases in Wuhan City, China. Imperial College London COVID-19 Response Team, **2020**. https://www.imperial.ac.uk/media/imperial-college/medicine/sph/ide/gida-fellowships/Imperial-College-COVID19-update-epidemic-size-22-01-2020.pdf

**Table 1.** Best-fit parameters of the SEIR model (2) for Hubei province and three cities in China.

| Place | $r$ | $r_2$ | $p$ | $p_2$ | $\alpha$ | $\gamma_\beta(t)$ |
|---|---|---|---|---|---|---|
| Hubei province | 10 | 10 | 0.035 | 0.030 | 1/7 | $0.05t^{-0.95}$ |
| Wenzhou | 15 | 15 | 0.065 | 0.065 | 1/7 | n/a [1] |
| Shenzhen | 10 | 15 | 0.045 | 0.045 | 1/7 | n/a |
| Shanghai | 15 | 15 | 0.065 | 0.065 | 1/7 | n/a |

[1] There was not enough reported data to reliably fit this parameter for this city.

**Table 2.** Initial values needed for the SEIR model (2) for Hubei province and three cities in China.

| Place | $S(0)$ | $E(0)$ | $I(0)$ | $R(0)$ | $D(0)$ |
|---|---|---|---|---|---|
| Hubei province | 61,499 | 4007 | 494 | 31 | 24 |
| Wenzhou | 287 | 223 | 0 | 2 | 0 |
| Shenzhen | 338 | 69 | 13 | 2 | 0 |
| Shanghai | 164 | 126 | 50 | 3 | 0 |

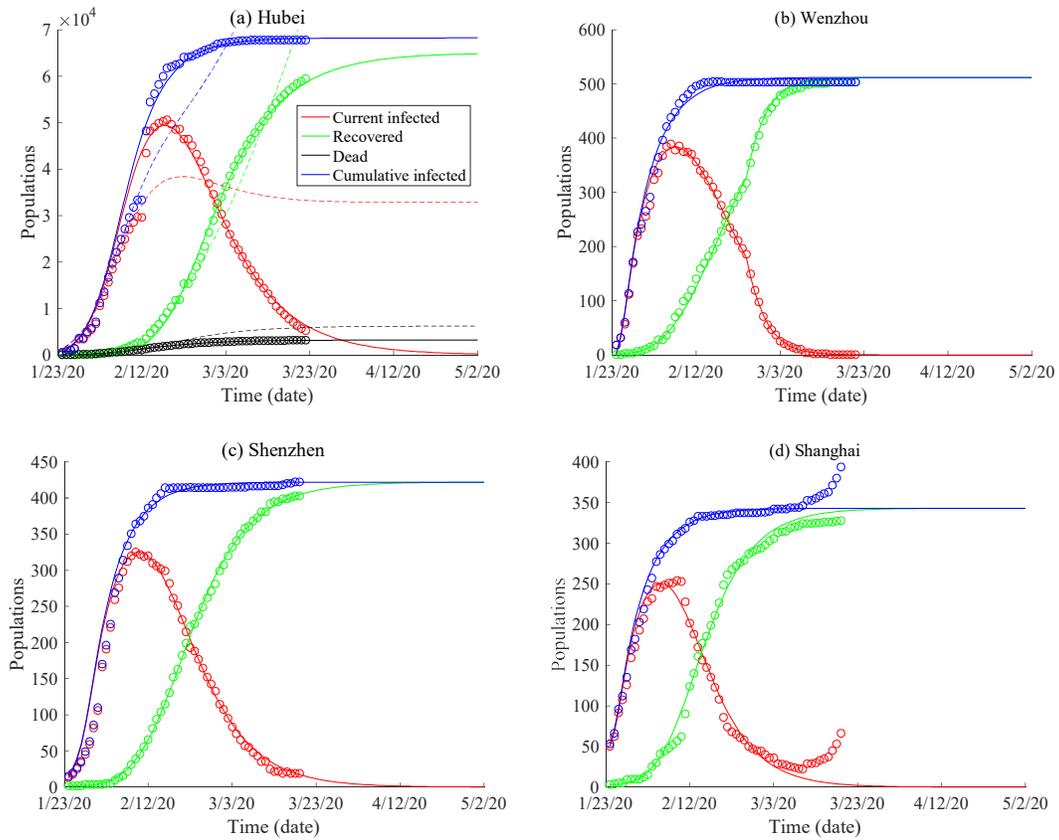

**Figure 1. SEIR model results**: the observed infection and recovery curves (symbols) versus the solutions of the SEIR model (2) (solid lines) for (a) Hubei province, and the cities of (b) Wenzhou, (c) Shenzhen, and (d) Shanghai in China. In (a), the blue and red dotted lines represent solutions of the constant-parameter SEIR model, and the black dots represent the death toll simulation using the classical PDE $dD/dt = \gamma_\beta I$.

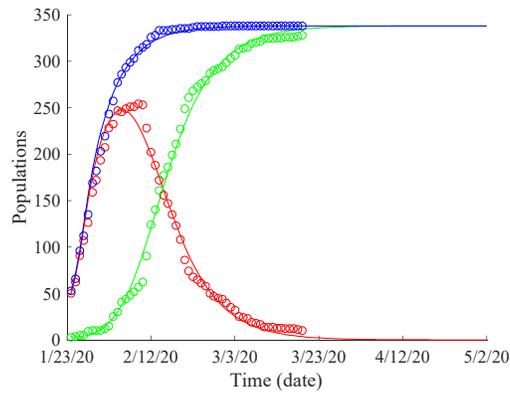

**Figure 2. SEIR model results**: the observed COVID-19 spread data (symbols) versus the solutions of model (2) (solid lines) for Shanghai City, China, after deleting the coronavirus cases imported from overseas.

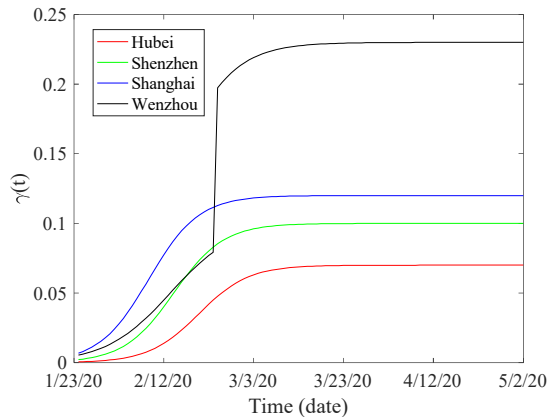

**Figure 3.** Comparison of the time-dependent recovery rate $\gamma(t)$ fitted for the four areas in China including Hubei province and the cities of Shenzhen, Wenzhou, and Shanghai.

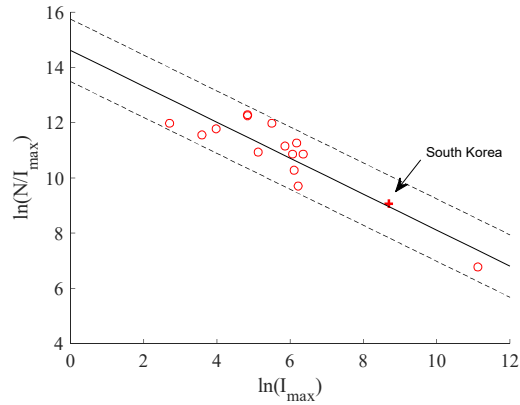

**Figure 4.** Index of infection severity ($C$) versus the maximum cumulative number of infections ($I_{max}$) calculated for different places in China (circles) and South Korea (the plus symbol). The best-fit relationship (shown by the solid line) is $N = e^{14.62} I_{max}^{0.349}$. The dashed lines represent the 95% confidence bounds.

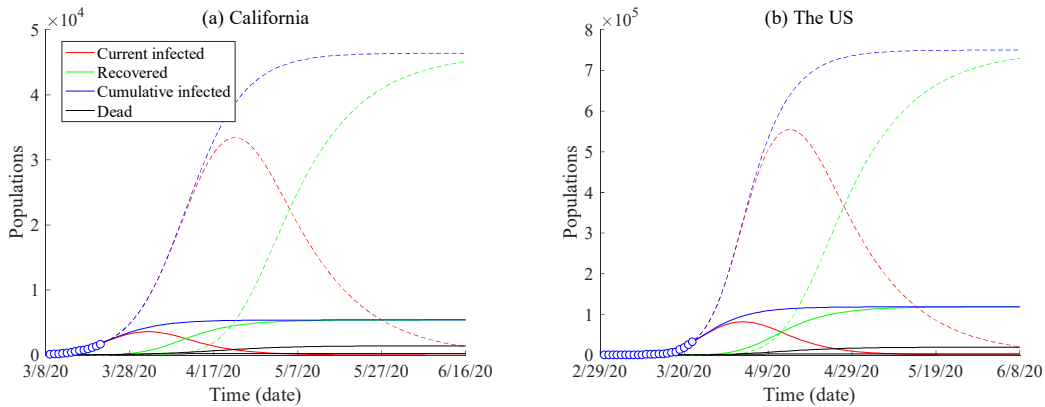

**Figure 5. SEIR model results**: Predictions of the time series of the infected and recovered population of COVID-19 in (a) California, U.S., and (b) the U.S. The solid lines represent the scenario with virus propagating trend similar to Shanghai, China; and the dashed lines follow the trend in Hubei, China.

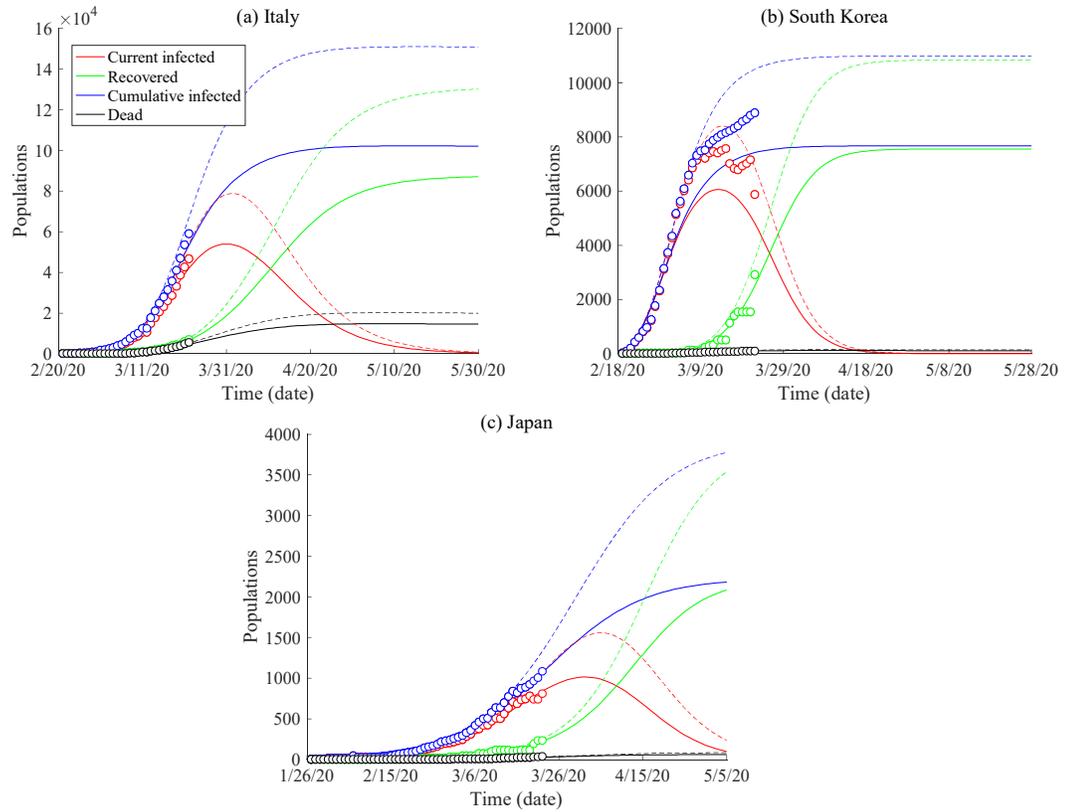

**Figure 6. SEIR model results**: Prediction of COVID-19 spread in different regions in the world (with the measured data up to 3/22/2020): (a) Italy, (b) South Korea, and (c) Japan. Data sources are from [1].

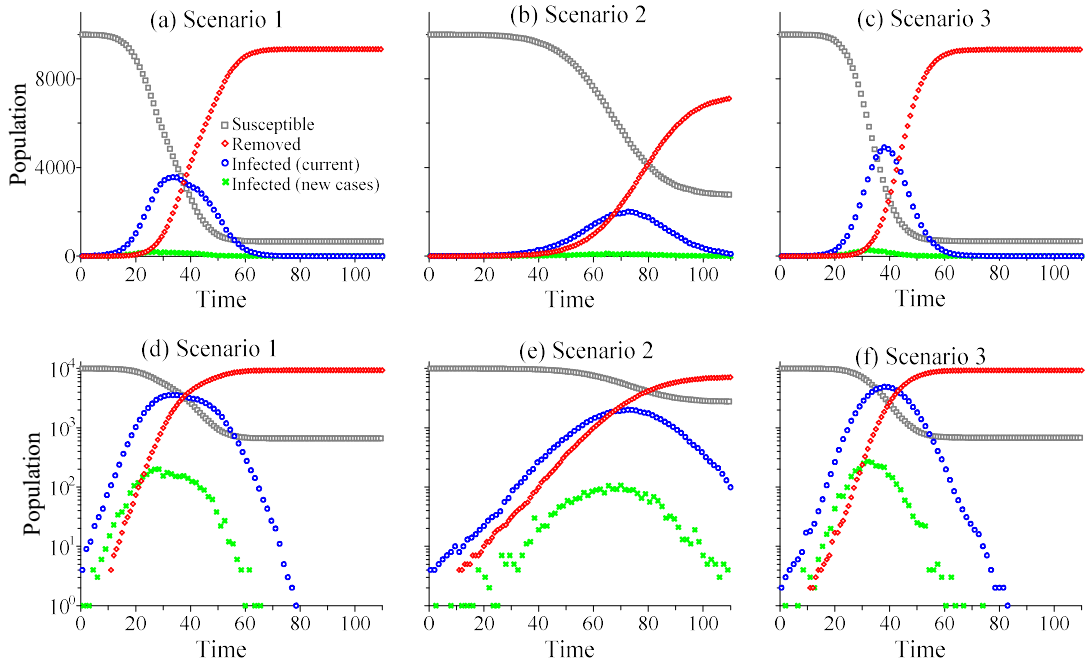

**Figure 7. Stochastic modeling**: the simulated virus spread in the city using the Lagrangian approach for the three mitigation scenarios: Scenario 1 (a), Scenario 2 (b), and Scenario 3 (c). The bottom plot is the semi-log version of the top plot, to show the tails of each curve.

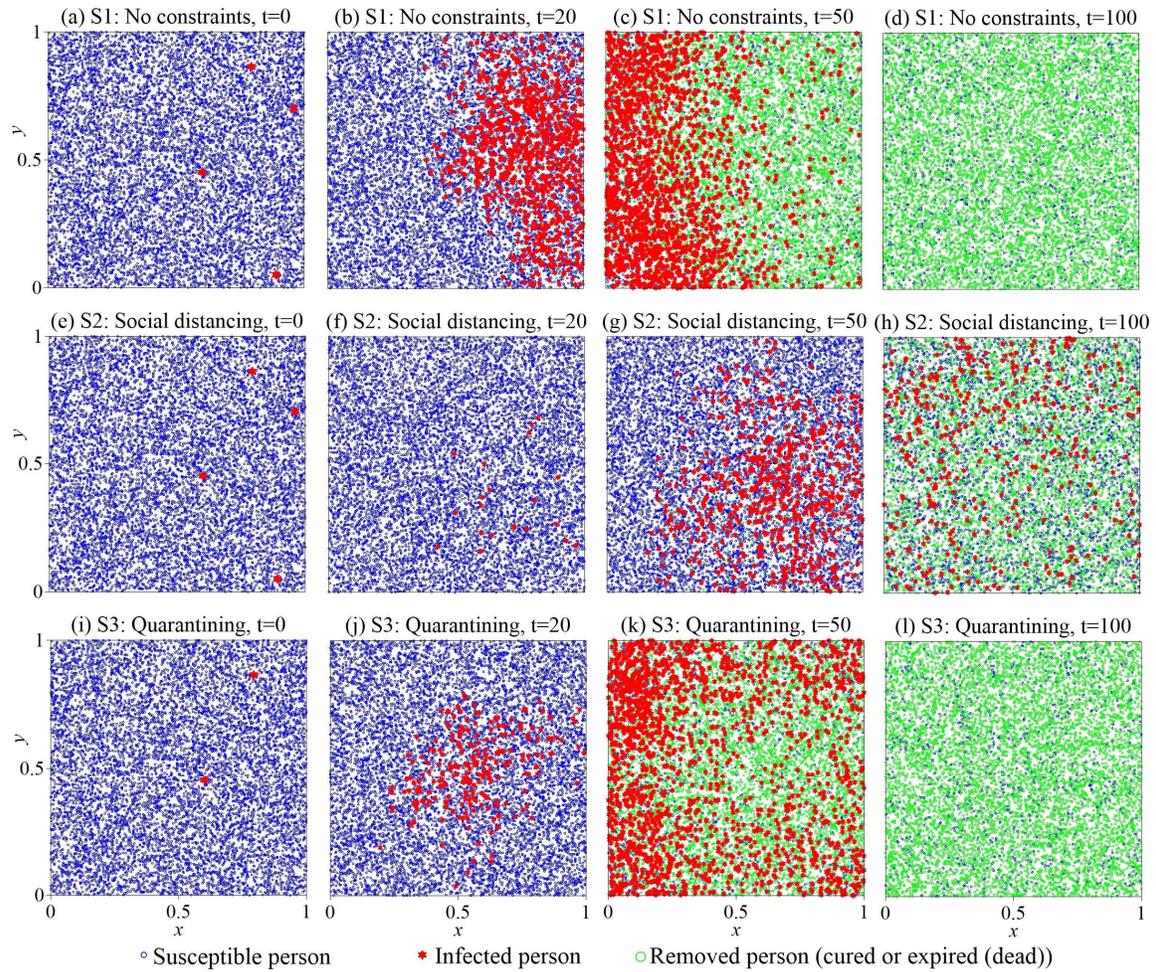

**Figure 8. Stochastic modeling**: the simulated particle plumes at different times (t=0, 20, 50, and 100 at the 1st, 2nd, 3rd, and 4th columns) for three scenarios to mitigate the COVID-19 outbreak in a city: Scenario 1 (S1): no constraint (1st row) (a)~(d), Scenario 2 (S2): Social distancing (2nd row) (e)~(h), and Scenario 3 (S3): Quarantining (3rd row) (i)~(l).